\documentstyle{article}
\begin{document}
\mbox{~}
\vspace{2cm}

\begin{center}
{\Large\bf Random Graph Gauge Theories as Toy Models\\[0.2cm]
for Non-perturbative String Theories}\\[1.0cm]
{\Large Thomas Filk}\\[0.3cm]
{\large University of Freiburg\\
Department of Physics\\
Hermann-Herder-Str.\ 3\\
D-79104 Freiburg\\
Germany}\\[0.2cm]
E-mail: \verb+thomas.filk@t-online.de+
\end{center}
\thispagestyle{empty}
\vspace{2cm}

\begin{center}
{\bf Abstract}
\end{center}
\vspace{0.2cm}

We present simple models which exhibit some of the remarkable features
expected to hold for the yet unknown non-perturbative formulation of 
string theories. 
Among these are (1) the absence of a background or embedding space for the
full theory, (2) perturbative ground states (local minima of the action)
having the characteristics of spaces of different 
dimension, (3) duality transformations between large and small coupling
expansions, and (4) perturbative excitations of
these ground states which can be interpretated as
string world sheets or p-brane world volumes. 
In this context we formulate gauge theories on arbitrary graphs and
speculate about actions for graphs which in a continuum and/or thermodynamic 
limit might be related to the Einstein-Hilbert action.
\vspace{1cm}
\begin{flushright}
Freiburg-preprint 00/10\\
July 2000\\
\end{flushright}
\newpage

\section{Introduction}
\setcounter{page}{1}

One of the most striking aspects of M-theory is that it seems to have
perturbative regimes with background spaces of different dimensions and
geometrically different objects (strings, p-branes etc.) as the 
corresponding elementary excitations (see e.g.\ \cite{Polchinski}). 
One way to relate spaces of different dimensions is
t-duality: let one of the dimensions of the background space be 
compactified to a cylinder of radius $R$ then in the presence of strings 
there exist duality relations
to a model with a background space having a compactified dimension being 
a cylinder of radius $1/R$. Therefore, in the
limit $R\rightarrow \infty$ (and may be some tuning of other couplings) one
finds a duality relation between background spaces of different dimension.
However, not all relations between different regimes and dimensions can be 
understood this way. So it may be helpful to look for objects which 
include regular ``spaces'' of given dimensions as special cases, but 
which are more general. Graphs are one example for such objects.  

Graphs have a relatively simple structure (they are sets with a symmetric 
relation, details will be explaind in sec.\ \ref{sec1}) 
but they allow for more complex derivatives. 
Any regular lattice\footnote{Lattices are usually defined as invariant 
representation spaces of some group of discrete translations (up to
boundary conditions) 
and, therefore, by definition
regular. However, in the context of discretized space-time structures
one sometimes reads about ``random'' lattices, meaning combinatorical
triangulations or cell complexes. For this reason, we emphasise at this
point the
regularity of a lattice. In the following, lattices will always be
assumed to be regular, otherwise we will speak of graphs.}, 
for instance, can be
considered as a graph. Lattices serve as background
``spaces'' in lattice models like lattice gauge theories. Therefore, although
graphs may be defined without reference to a
background space or a dimension, they can dynamically provide
background spaces with a fixed dimension. So one might wonder, 
if there exist actions or weights on the set of all graphs 
such that lattices (amongst others) are the local minima of these 
actions. Or, to be more speculative and ambitious,
does there exist an ``Einstein'' action on the set
of graphs such that the local minima correspond to discretized solutions
of Einsteins equations in different dimensions.
In this case the local minima would provide the background spaces, 
and perturbations around these minima (defects like dislocations or
disclinations etc.) might be considered
as fluctuations of this background space with the physical interpretation
depending on the geometric structures of these fluctuations. The appealing
feature of this picture is that background space and excitations are
the same objects - parts of a graph. This reminds us that in general 
relativity the splitting of the metric for perturbative expansions,
$g_{\mu \nu}=\eta_{\mu \nu}+h_{\mu \nu}$, introduces two objects which
seem to be of completely different natur in the perturbative picture
(a static background space and a particle propagating in this space), but
are in principle the same objects on a non-perturbative level. 

Another striking feature of some of the perturbative regimes of M-theory is 
s-duality: the perturbation 
expansion in some coupling $g$ is related to some other perturbation
expansion in $1/g$ (or, more general, expansions for small $g$ are related
to expansions for large $g$). This kind of duality has a long history in
statistical mechanics, the most famous example being the self-duality of
the Ising model on a 2-dimensional square lattice (\cite{Kramers},
see e.g.\ \cite{Feynman,Baxter}). In 1971 Wegner
\cite{Wegner} found more general duality relations among a whole class
of spin models including spin-gauge theories on lattices of higher
dimension. The first example of duality in quantum field theory was the
Coleman-Mandelstam duality between the Sine-Gordon theory and the
massive Thirring model in 2 space-time dimensions 
\cite{Coleman, Mandelstam}. In this case the dual fields could be
constructed explicitly. In later generalizations of duality to non-abelian
gauge theories \cite{Olive} this was not possible, although there was a
general believe that the ``defects'' of one model are related to the 
fundamental excitations of the dual model. More recently it is the 
AdS-CFT duality (see, e.g., \cite{Maldacena} and references therein), which 
seems to be of direct relevance to string theory.

In this paper we consider the spin models and spin-gauge theories of
Wegner as the most simple examples of models with duality. But,
why might gauge theories be appealing candidates in the search for M-theory?
Usually
the argumentation goes the other way round: string theory leads naturally
to gauge theories as well as to general relativity, because
(super)-string theory includes massless spin-1 and spin-2
particles in a covariant way. But the argument may be turned around, at 
least partly. Gauge theories (most explicitly on the lattice, see e.g.\ 
\cite{Rothe} and references therein) allow for a strong 
coupling expansion in terms of closed surfaces - ``string world-sheets''.
Apart from free gauge theories (like electrodynamics)
a whole spectrum of particles - glueballs - can be obtained from loop-loop
correlation functions, including spin-1 and spin-2 particles. These
particles, however, are massive and therefore no
candidates for a gauge theory or general relativity. 
Another difference is
that the expansion parameter in string theory is
the weight for string sheet topologies while in gauge theories the
weight for closed surfaces depends on their area and, possibly (depending 
the gauge group), on local intersections of surfaces. It is not clear how
to formulate a gauge theory such that in the large coupling expansion the 
weight for the surface topologies can be tuned. Nevertheless, a non-trivial
gauge theory with massless spin-1 and spin-2 ``glueballs'' might be an
interesting alternative to (or candidate for) non-perturbative string
theories. These remarks merely serve 
as a motivation for studying gauge theories in the
context of non-perturbative string theory.
\vspace{0.2cm}
     
The models I shall discuss in this paper exhibit some of the
remarkable features expected for the non-perturbative formulation of
string theories, among which are:
\begin{enumerate}
\item
The formulation of these models make no explicit reference to a
background or embedding space, a dimension, or a geometric structure
(apart from the structures inherent to graphs).
\item
Depending on the values of the coupling constants, the
actions for these models have enumerous local minima, some of which
correspond to lattices with different dimensions for
the different ground states.
\item
Not all local minima of the action correspond to flat spaces but also other 
forms of ``background spaces'' - e.g., compactified dimensions - exist.
\item
In some cases, perturbations around these ground states can be
interpreted as summations of closed surfaces, the euclidean equivalent
of world sheets of strings, or closed volumes of higher dimension, the
analoga of world volumes of $p$-branes. 
\item
In some cases duality transformations exist which transform
the perturbation expansion for a small coupling constant into a
different perturbative regime where this coupling is large.
\end{enumerate}
In this paper I shall mainly concentrate on two special models: 
${\bf Z}_2$-spin (Ising) model and ${\bf Z}_2$-gauge theory coupled to 
random, non-directed and simple graphs. The generalization of
these models to arbitrary scalar fields or 
gauge groups is straight forward and will be
sketched. However, the implementation of duality transformations for these 
generalized 
cases (especially for non-abelian gauge groups) is more difficult.

In section \ref{sec1}, we review some simple structures of graph theory, 
formlate actions for graphs (weights on the set of graphs), and discuss
their properties.  

Section \ref{sec2} will summarize the formulation of ${\bf Z}_2$-spin,
${\bf Z}_2$ gauge, and ${\bf Z}_2$ $p$-gauge (analoga of gauge theories of
$p$-forms) theories on hypercubic (regular) lattices. 
In this context, we also review the
duality transformations among these models, first formulated by Wegner, 
as relations between the small and high temperature perturbation 
expansions, which can be interpreted as summations
over objects of different geometrical structure.

In section \ref{sec3}, scalar and gauge field theories are formulated on
arbitrary graphs, and the features of the ${\bf Z}_2$ cases are discussed 
in more detail. Finally, in sec.\ \ref{sec4}, I will speculate about further 
generalizations. 

\section{Random Graphs}
\label{sec1}

In this section, I shall first review some of the algebraic properties of
graph theory, in particular the notion of the adjacency matrix and related
structures (Sect.\ \ref{sec11}). Next, I will formulate different 
contributions to an action for random graphs and discuss their properties
(Sect.\ \ref{sec12}). I will also speculate 
about an action
for graphs which might be related to the Einstein-Hilbert action in
general relativity (Sect.\ \ref{sec13}).

\subsection{The adjacency matrix}
\label{sec11}

The following notions can be found in any good book on graph theory, e.g.\
in \cite{Biggs} where the algebraic properties of graphs are emphasized.

We will only consider undirected graphs without self-loops and multiple
connections between vertices. In this case, a graph can be defined as
a set $V$ (the set of vertices) together with a symmetric, non-reflexive
relation stating which of the vertices are neigboured. The
adjacency matrix $A$ is the matrix of this relation, i.e.
\begin{equation}
        A_{ij} ~=~ \left\{ \begin{array}{ll}
             1 & \mbox{if $i$ and $j$ are neighbors}\\ 0 & 
                {\rm otherwise}~.
             \end{array} \right.   
\end{equation}
$A$ is symmetric and its diagonal elements are 0. Two graphs are considered
to be equivalent, if they differ only by a relabeling of their vertices.
Therefore, two adjacency matrices $A_1$ and $A_2$ correspond to the same
graph, if
\[        A_2 ~=~ P^+ A_1 P   \,   \]
where $P$ is some permutation matix, acting on the functions over the set of 
vertices. If, for an adjacency matrix $A$, we have
\begin{equation}       
                  A ~=~ P^+AP  \hspace{0.7cm} {\rm or}
                 \hspace{0.3cm} [A,P] ~=~ 0  
\end{equation}
for some permutation $P$, we call $P$ a symmetry of $A$. The set of all
$P$'s which are symmetries of $A$ generate the symmetry group of that 
graph. (The size of this group determines the symmetry factors for
Feynman diagrams in a QFT perturbation expansion.)       

We define a {\em L-path} on a graph from a vertex $i$ to a vertex $j$ as a
sequence of vertices $i=k_0,k_1,k_2,...,k_{L-1},j=k_L$, such that
$k_n$ and $k_{n+1}$ are neigbored vertices, i.e.\ $A_{k_n k_{n+1}}=1$, for
all $n=0,...,L-1$. The number
of steps, $L$, is called the length of the path. If all vertices of a
graph can be connected by paths, we call the graph connected.

The adjacency matrix
$A$ may also be considered as the generator of the number of random paths
on the graph:
\[  (A^L)_{ij} ~=~ \mbox{number of paths from $j$ to $i$ of length $L$}
   ~=~ \mbox{number of $L$-paths from $j$ to $i$} \,.\]
This follows immediately from the definition of $A$ and the definition of
paths, as
\[  (A^L)_{ij}~=~ \sum_{k_1} \sum_{k_2}\cdots \sum_{k_{L-1}}
              A_{ik_1}A_{k_1 k_2} \cdots A_{k_{L-1}j}    \, . \]  
The only non-zero contributions to this expression come from sequences for
which the elements of the adjacency matrix are all 1, i.e.\ paths from $j$
to $i$. In particular,
\[       (A^2)_{ii} ~=~ \delta_i ~=~ 
               \mbox{number of incident lines to $i$} \, , \]
i.e.\ the {\em degree} $\delta_i$ of $i$, and
\[       {\rm tr}(A^2) ~=~  2 E \, ,  \]
where $E$ is equal to the total number of edges (lines) of the graph. (We
mostly consider graphs with a finite number of vertices and lines. 
However, in
statistical mechanics one is usually interested in the thermodynamic limit
where $V$ and $E$ become infinite. The problems related to this limit
are well known shall not be discussed here.) For later purposes, we
also define the {\em valence matrix} $V$ of the graph as the diagonal
matrix of the degrees of the vertices:
\[           V_{ij} ~=~ {\rm diag} \{ \delta_i \}  \, . \]

If all vertices of a graph have the same degree $\delta$, the graph is called
regular of degree $\delta$. Let $K_{ij}=1$, then
\[         [K,A] ~=~ 0 ~~\mbox{iff $A$ is the adjacency matrix of a 
            regular graph.} \]
The proof is straight forward (see, e.g., \cite{Biggs}). 

The geodetic distance $r(i,j)$ between two vertices $i$ and $j$ on a graph is
defined by the length of the shortest path connecting
these two vertices. The number of vertices within a distance $r$ from
a given vertex $i$ is called the volume of a ball of radius $r$ around
$i$: 
\[   {\rm Vol}_i(r)= | \{ j | r(i,j) \leq r \} |  \, . \]
For infinite graphs, this allows the definition of a dimension $d$, 
\begin{equation}
\label{eqHausdorff}
      d ~=~ {\rm inf} \left\{ x | \lim_{r \rightarrow \infty}
            {\rm Vol}(r) / r^x ~<~ \infty \right\}  \, , 
\end{equation} 
which is often formulated in terms of 
the following scaling relation
\begin{equation}
\label{eqVol}
       {\rm Vol}(r) ~\stackrel{r\rightarrow \infty}{\longrightarrow}
                      ~ {\rm const}\; r^d(1+O(1/r)) \, ,   
\end{equation}        
although the existence of such a scaling relation is more restrictive.
For lattices this definition agrees with the usual notion of the
dimension of a lattice. This intrinsic definition of a dimension has long
been used in the theory of clusters or percolation models (see e.g.\
\cite{Stauffer}). A recent and more mathematical treatment can be found
in \cite{Requardt}.

Finally, I will discuss some properties of the graph Laplacian.
In order to motivate and justify this name, we consider a scalar
Gaussian field defined on the vertices of the graph. The natural action
is
\begin{eqnarray}
\label{eqpos} 
         S[A;\varphi] &=& \frac{1}{4}\sum_{ij} A_{ij} 
                                       (\varphi_i - \varphi_j)^2 \\
\nonumber       &=& \frac{1}{4} \sum_{ij}
           (2 A_{ij} \varphi_i^2 - 2 A_{ij} \varphi_i \varphi_j ) \\
       &=& - \frac{1}{2} \sum_{ij} \varphi_i \Delta_{ij} \varphi_j \, , 
\end{eqnarray}
where
\[       \Delta ~=~ A - V    \]
is the graph Laplacian.
$\Delta$ is a generalization of the usual concept
of a Laplacian to arbitrary graphs and is easily shown to be equivalent to
the second difference quotients summed over all directions on a 
lattice. (In the theory of electrical networks $-\Delta$ is sometimes
referred to as the matrix of admittance.)
As the action in eq.\ (\ref{eqpos}) is always non-negative, $-\Delta$ 
is also a non-negative matrix. The number of zero-modes is equal to
the number of components of the graph. Hence, for connected graphs the
graph Laplacian has exactly one zero mode. As usual
\[  \int \! \prod_i {\rm d}\varphi_i \;
          \exp - \left( \frac{1}{4} \sum_{ij} A_{ij}(\varphi_i - \varphi_j)^2
             + \frac{1}{2} \mu^2 \sum_i \varphi_i^2 \right)   ~\propto~
              \left( {\rm det}(V-A+\mu^2)\right)^{-1/2} \, . \]

\subsection{An action for random graphs}
\label{sec12}     

Next, I will write down an action on the set of graphs. For simplicity, I will
restrict myself to the set of all graphs with a fixed number of vertices.
It should be kept in mind, however, that the proposed
action is only one possible choice out of many. The aim is
to find an action for which some of the local minima may be interpreted as
background spaces (``flat''
lattices, or lattices with nontrivial compactifications
for some directions). For this purpose, other actions might be more 
suitable; the
one formulated in this section is just a candidate. 

The action consists of three terms related to the properties of the
adjacency matrix discussed in the previous subsection and is defined to be  
\[     S[A] ~=~ \alpha S_1[A] + \beta S_2[A;\delta] + S_3[A;\mu]   \, , \]
where
\begin{eqnarray}
\label{eqperm}     
            S_1[A] &=& \sum_P {\rm tr}(P^+AP-A)^2 \\
\label{eqdelta}
            S_2[A;\delta] &=& \frac{1}{2} \sum_i (\delta_i - \delta)^2  \\
            S_3[A;\mu] &=& \ln \mu - \frac{1}{2}\ln {\rm det}(V-A+\mu^2) \, .
\end{eqnarray}
I shall now discuss the meaning of the different terms. The parameters
$\alpha, \beta, \delta$ and $\mu$ might be tuned in order to enhance 
or suppress the contribution of graphs with certain properties.

\begin{enumerate}
\item
The summation runs over all permutations of vertices. Each term 
${\rm tr}(P^+AP-A)^2$ is a measure as to how far $A$ deviates from being 
symmetric under $P$. If $P$ is a symmetry of $A$, the term vanishes.
$S_1[A]$ is zero for two graphs: the empty graph (all points are isolated,
there are no lines) and the complete graph. These two are the absolute
minima of $S_1$. Local minima of $S_1$ should correspond to graphs
with a large symmetry. ``Local'' refers to the set of graphs and may
be made more explicit with a distance functional
\[    D(A_1,A_2) ~=~ {\rm min}_P\; {\rm tr}(A_1 - P^+A_2 P)^2  \, . \]
Formally, $S_1[A]$ resembles a nearest neighbor interaction of
matrix models. In this case,
\begin{equation}
\label{eqtrans}
      S_M[A] ~=~ \sum_{i=1}^d {\rm tr}(T_i^+AT_i-A)^2  \, , 
\end{equation}
where $T_i$ is the translation matrix for one step in direction $i$ and 
the sum runs over all directions. Equation 
(\ref{eqtrans}) is minimized for adjacency matrices $A$ having the symmetry 
of the translation group, i.e., $A$ corresponds to a lattice graph.
Translations form a subgroup of the permutation
group. However, using the full permutation group in eq.\ (\ref{eqperm})
has the advantage that no dimension is specified.
\item
$\delta_i$ is the degree of vertex $i$. Therefore, this term 
suppresses graphs for which the degree at the vertices deviates much
from a prefered value $\delta$. Noticing that
\[      \sum_i \delta_i^2 ~=~ {\rm tr}A^2K \hspace{1cm} {\rm and}
         \hspace{1cm} \sum_i \delta_i ~=~ {\rm tr}AK   \]
we can rewrite the second term also as
\[        S_2[A;\delta] ~=~ \frac{1}{2} {\rm tr} (A-\delta)^2K   \, . \]
We might also have added a term
\begin{equation}
\label{eqKkom}
                    S'_2[A] ~=~ {\rm tr} [K,A]^2   
\end{equation}
which suppresses all graphs which deviate from being regular without 
reference to a certain degree.

It might seem that specifying a degree $\delta$ implicitly specifies
a dimension for the graph. This is not true. For any given degree 
$\delta>2$, there exist graphs of any dimension. The two extreme cases are
represented, e.g., by the Bethe lattice (Cayley tree) giving $d=\infty$, 
and by a linear chain of complete graphs glued together appropriately giving
$d=1$.
\item  
The first two actions in combination will suppress graphs with very 
irregular degrees (like complete subgraphs embedded in an otherwise empty 
environment), or with degrees which deviate much from $\delta$,
and they will enhance graphs with high symmetry. There exist very disconnected
graphs which satisfy both properties. Consider, e.g., an ensemble of
disconnected complete graphs of degree $\delta$. They minimize action 2
and they also are local minima of action 1. The last action, $S_3[A]$, may 
be used to suppress graphs with more than one component.
In the limit $\mu \rightarrow 0$ this term becomes infinite unless
the graph is connected.

The determinant of the graph Laplacian may also be used to
enhance regular graphs with many small loops \cite{Ambjorn}, again
supporting lattices.  
\end{enumerate} 
    
\subsection{An Einstein action for graphs?}
\label{sec13}

The actions mentioned in the previous section enhance graphs which have a
large symmetry, small fluctuations in the valencies, and which are
connected. It is not expected that a combination of these terms 
leads to an action which resembles Einsteins action or which has the
same local minima as Einsteins action (not even in a certain limit).
Therefore, the question arises, if it is possible to formulate an
Einstein action for arbitrary graphs which does not
refer to a definite dimension. Minima of this action would correspond
to solutions of Einsteins equations in different space-time dimensions.
  
Discrete Einstein actions have been used in Regge calculus \cite{Regge}
and for combinatorical triangulations \cite{Ambjorn2}. However, in both 
cases the dimension is explicitly given, in the first case even a
metric structure. In the second case curvature is ``counted'' by the
numbers of cells meeting at subcells of codimension 2 \cite{Ambjorn2}.
Although the actions (\ref{eqdelta}) and (\ref{eqKkom}) look similar,
they are not supposed to be real analogues of an Einstein action.
Only if we require additional structures for the graphs (for instance
``planarity'', as in the case of 2-dimensional triangulartions and matrix
models \cite{Ambjorn2}) a relation to Einsteins action might be proven.  

Several approaches come to ones mind to find an analoga of Einsteins action
for arbitrary graphs. One is related to the following
relation known from differential
geometry (see e.g.\ \cite{Pauli}):
\begin{equation}
\label{eqPauli}
   {\rm Vol}_x(r) ~=~  c_1 r^d \left( 1 - c_2 r^2 R(x) + 
                                                       O(r^3) \right)\, . 
\end{equation}
On the left hand side, ${\rm Vol}_x(r)$ denotes the volume of a ball of 
radius $r$ around point $x$. 
The leading term on the right hand side ($c_1$ and $c_2$ are positive 
constants) is just the volume
of a ball in $d$-dimensional flat space, the first correction term
contains the scalar curvature $R(x)$ at point $x$. However, to apply this 
formula to graphs one has to deal with the following problem first. 
Equation (\ref{eqPauli})
holds in the limit $r \rightarrow 0$, while scaling relations of the
type (\ref{eqVol}) hold for the limit $L \rightarrow \infty$. This is 
related to the fact that in differential geometry manifolds are supposed
to approach flat spaces when viewed at sufficiently small regions, while 
random graphs are expected to approach homogeneous (not necessarily flat)
structures when viewed at large scales. Therefore, the best one can hope
for is that for graphs a relation like (\ref{eqPauli}) only holds in a
``scaling window'', i.e., for $r$ not too small (where the discrete structure
of the graph dominates) and not too large (where higher order corrections
to eq.\ (\ref{eqPauli}) are important).   

A second approach is based on the following asymptotic expansion for
the heat kernel of the Laplace operator $\Delta[g]$ on a manifold of
dimension $d$ with
metric $g$ (for simplicity we assume that the manifold is compact and has 
no boundary) 
\cite{Gilkey,Eguchi}:
\begin{equation}
\label{eqheat}
    {\rm tr~e}^{\Delta[g] t} ~ \stackrel{t \rightarrow 0}{\longrightarrow}
         ~ c'_1 \frac{{\rm Vol}}{t^{d'/2}} - 
          \frac{c'_2}{t^{(d'-2)/2}} \int \sqrt{g} R + \cdots  \, .
\end{equation}
The leading term is simply the total volume of the manifold and corresponds 
to a cosmological term in the Einstein-Hilbert action. (Note that the
spectral dimension $d'$ in eq.\ (\ref{eqheat}) may differ from the
Hausdorff dimension $d$ in eq.\ (\ref{eqPauli}). Requiring them to be
equal again suppresses ``irregular'' graphs.) 
The first correction is the integrated scalar
curvature, i.e., the Einstein term. Therefore, it is conceivable to use
the asymptotic behavior of the spectral density of the graph Laplacian
for a construction of an action related to the Einstein-Hilbert action.
Much is known about the spectrum
of the adjacency matrix and the graph Laplacian (see e.g.\ \cite{Doob}),
so in this case even analytical investigations might be possible.     

A third approach is closely related to the previous one. Let $(A^L)_{ii}$
be the number of paths of length $L$ starting and ending at vertex $i$, and
let $N_i(L)=\sum_j(A^L)_{ij}$ be the total number of paths of length $L$
starting at $i$. Then
\[       R_i(L) ~=~ \frac{(A^L)_{ii}}{N_i(L)} ~
         \stackrel{L\rightarrow \infty}{\longrightarrow} ~
        \frac{c}{L^{d/2}} + \cdots   \]
defines a recurrence probability of paths of length $L$. The leading term
is again given by the volume and dimension of a ball around $i$ on the
graph and the corrections are again expected to be related to the scalar 
curvature. The analogy with the previous approach will be obvious if one
interpretes the diagonal elements of the heat kernel, 
$({\rm e}^{\Delta[g]t})_{ii}$, as the recurrence propabilities of a
diffusion process on the graph.

These are only a few examples for expressions which for regular manifolds
are known to have an expansion where the leading terms are related to
the volume(density) and the dimension of the manifold and the correction
terms are related to the scalar curvature, and which all have analogues
also for arbitrary graphs. Therefore, expressions of this type have a
good chance to define an action for graphs which in a sensible
continuum and termodynamic limit give rise to a cosmological term and
an Einstein action. However, much analytical and, presumably, even more
numerical work will be necessary until an adequate action for graphs will 
be found.
 
\section{$Z_2$-models on hypercubic lattices}
\label{sec2}

The ${\bf Z}_2$-spin model and the ${\bf Z}_2$-gauge theory 
are the simplest examples among a class of models discussed
intensively by Wegner \cite{Wegner}. Of special relevance are the different
duality relations among these models on $d$-dimensional lattices. In this
section we will restrict ourself to the self-dual hypercubic lattices.

Let us consider ${\bf Z}_2$ variables ($z\in \{+1,-1\}$) associated to the
$p$-dimensional elements (cells) of a hypercubic lattice.
These are the vertices ($p=0$), the lines ($p=1$), the plaquettes ($p=2$),
the cubes ($p=3$), etc. The $p+1$-dimensional cells are bounded by
$2(p+1)$ $p$-cells. Define the action to be
\[  S_p ~=~ - \sum_{(p+1)-{\rm cells}} 
           \left( \prod_{\partial (p+1)-{\rm cell}} z_p \right)  \, , \]
where the summation extends over all $(p+1)$-cells of the lattice and for 
each cell we take the product of spin variables on its boundary.
The partition function is given by
\[   Z_p ~=~ \sum_{\{z\}} {\rm e}^{-\beta S_p} \, . \]
For $p=0$ we obtain the Ising model with the spins $z$ attached to the
vertices of the lattice and the action given by
\[        S_0[z] ~=~ - \sum_{\langle i,j \rangle} z_i z_j   \, . \]
The summation extends over all links (cells of dimension 1) of 
neigbored vertices $i$ and $j$.

The case $p=1$ yields the standard ${\bf Z}_2$-gauge theory with the group
variables $z=\pm 1$ associated to the links and the action given by
\[   S_1[z] ~=~ - \sum_{p_2\simeq \{ijkl\}} z_{ij} z_{jk} z_{kl} z_{li} \, ,\]
where now the sum extends over all plaquettes (cells of dimension 2) of the 
lattice and for each
plaquette we take the product of the four spin variables on the links
which form the boundary of this plaquette.

In the case $p=2$ the spin-variables are associated to the plaquettes of
the lattice, the action consists of a sum over all cubes and for each cube
one takes the product of the spin-variables on the six plaquettes which
form the boundary of that cube. All models for 
$p\geq 1$ have a gauge invariance. As the
fundamental degrees of freedom are attached to the $p$-cells
of the lattice, they may be considered as lattice analoga of gauge 
theories of $p$-forms.

For all models there exists a high temperature expansion and a low 
temperature expansion similar to the one known for the Ising model
(see e.g.\ \cite{Feynman,Baxter}), the difference only being the geometrical 
objects on the lattice one has to sum over. In all cases the high temperature 
expansion (small $\beta$) has the following form 
\[    Z_p ~=~ 2^{E_p} (\cosh \beta)^{E_{p+1}}
        \sum_{L=0}^\infty t_p(L) (\tanh \beta)^L   \, , \]
where $E_p$ and $E_{p+1}$ are the total number of $p$ and $p+1$-cells,
respectively, and
$t_p(L)$ a combinatorical factor whose geometric interpretation depends on
$p$. For $p=0$ (the Ising model), $t_0(L)$ 
denotes the number of closed polygones of length $L$ on the lattice. 
In this case, the high temperature 
expansion may be interpreted as a summation over all closed polygones on 
the lattice where each polygone is weighted by a factor depending only on 
its length (the number of links).  

For the ${\bf Z}_2$-gauge theory, $t_1(L)$ equals
the total number of closed surfaces (obtained by pasting together plaquettes)
of area $L$ (number of plaquettes) on the lattice. Similarly for the other
models: The high temperature expansion of the theory with variables defined
on the $p$-dimensional cells of the lattice consists of a summation over
all closed $(p+1)$-volumes obtained by pasting together $(p+1)$-dimensional
cells of the lattice, and the weight for each such volume depends only on
the number of the $(p+1)$-cells it contains. In the same way as the high
temperature expansion (large coupling expansion) of a gauge theory
is related to strings (summation over string world sheets) the high
temperature expansions of Wegners $p$-models are related to $p$-branes in
the sense that they generate a summation over world volumes of $p$-branes.

All models also have a low temperature expansion (large $\beta$) which reads
\[ Z_p~=~ 2\, {\rm e}^{\beta E_{p+1}} 
          \sum_{L=0}^\infty c_p(L)\, {\rm e}^{-2\beta L} \,.\]
Again these expansions only differ in the geometrical interpretation of
the combinatorical factors $c_p(L)$ (and the exponent of the overall factor).
$c_p(L)$ is equal to the total number of closed $(d-p-1)$-volumes on the
lattice. For the Ising model and $d=2$ these are again polygones showing 
the self-duality of the Ising model for $d=2$. For the Ising model on a
$d=3$ lattice these are closed surfaces. This
expresses the fact that the Ising model and the ${\bf Z}_2$-gauge theory
are dual to each other for $d=3$.  

This structure extends to the models where the ${\bf Z}_2$-variables are 
defined on $p$-cells, i.e.\ $p$-form gauge theories. 
The low temperature expansion contains a summation
over closed objects of dimension $d-(p+1)$, the high temperature expansion
a summation over closed objects of dimension $p+1$. 
Two models are dual to each other whenever the dimensions of these objects
match, i.e., when 
the high and low temperature expansion of one model coincides with
the low and high temperature expansion of the other model, respectively.
In this way we obtain a whole series of duality relations depending on
the value for $p$ and the dimension of the lattice. For the lowest
dimensions the duality relations are listed in table (\ref{tab1}).

\begin{table}[htb]
\begin{tabular}{|cccc|} \hline
\mbox{~} & & & \\[-0.2cm]
$d$ & duality relation & geometric duality 
                     & excitation duality \\[0.2cm] \hline
\mbox{~} & & & \\[-0.2cm]
2 & Ising $\leftrightarrow$ Ising & $\sum$ polygons $\simeq$ 
      $\sum$ polygons & particle $\leftrightarrow$ particle \\[0.2cm] \hline
\mbox{~} & & & \\[-0.2cm]
3 & Ising $\leftrightarrow$ 1-form gauge & $\sum$ polygons $\simeq$ 
      $\sum$ surfaces & particle $\leftrightarrow$ string \\[0.2cm] \hline
\mbox{~} & & & \\[-0.2cm]
4 & Ising $\leftrightarrow$ 2-form gauge & 
     $\sum$ polygons $\simeq$ $\sum$ 3-volumes & 
             particle $\leftrightarrow$ 2-brane \\[0.2cm]
4 & 1-form gauge $\leftrightarrow$ 1-form gauge &
     $\sum$ surfaces $\simeq$ $\sum$ surfaces & 
              string $\leftrightarrow$ string \\[0.2cm] \hline
\mbox{~} & & & \\[-0.2cm]
5 & Ising $\leftrightarrow$ 3-form gauge & 
     $\sum$ polygons $\simeq$ $\sum$ 4-volumes & 
        particle $\leftrightarrow$ 3-brane \\[0.2cm]
5 & 1-form gauge $\leftrightarrow$ 2-form gauge & 
     $\sum$ surface$\simeq$ $\sum$ 3-volumes &
       string $\leftrightarrow$ 2-brane \\[0.2cm] \hline
\end{tabular}
\caption{\label{tab1}%
Duality relations for the ${\bf Z}_2$ models on $p$-cells.}
\end{table}

\section{${\bf Z}_2$-Gauge Theory on Random Graphs}
\label{sec3}

Now we want to combine the models for random graphs and gauge theories.
Random graphs may yield lattices as local minima although no 
prefered dimension is put into the theory beforehand. Therefore, they
dynamically provide background spaces for the perturbative regimes. 
The spin models satisfy duality relations and have perturbation expansions
in terms of closed objects - lines, surfaces, etc.

Most of the models mentioned in the previous section have no 
natural generalization for arbitrary graphs because there are no
natural analogues for $p$-cells apart from $p=0$ (the vertices) and
$p=1$ (the lines). So, at first sight, only the Ising model seems to
have a natural formulation on graphs, its partition function 
defined by
\[        Z~=~ \sum_{\{z\}} \exp \left( \frac{1}{2}\beta \sum_{ij} A_{ij}
                   z_i z_j \right)  \, . \]
The high and low temperature expansions are unchanged:
\begin{eqnarray*}
 Z_0 &=& 2^{E_0}(\cosh \beta)^{E_1}\sum_{L=0}^\infty t_0(L) (\tanh \beta)^L \\
 Z_0 &=& 2\,{\rm e}^{\beta E_1} \sum_{L=0}^\infty c_0(L){\rm e}^{-2\beta L}\,, 
\end{eqnarray*}  
where now $t_0(L)$ and $c_0(L)$ are the number of tiesets and cutsets 
of length $L$ of the graph, respectively. Tiesets are again closed
polygones, or, to be more precise, subgraphs (subsets of lines together with
their incident vertices) where all vertices have even degree. 
To obtain a cutset one first chooses an arbitrary partition  
of the sets of vertices into two disjoint subsets (the ``spin-up''-set and 
the ``spin-down''-set). The corresponding cutset then consists of all 
lines of the graph which connect vertices of one set with vertices
of the other. On a hypercubic lattice the cutsets
are in one-to-one correspondence to the $d-1$-dimensional closed volumes on
the dual lattice.
 
The equivalence of these two expansions
provides an easy way to relate the number of cutsets to the number
of tiesets and vice versa, showing, e.g., that one set is fixed once the
other is given. But in general there will be no ``dual'' graph,
for which the tiesets are equivalent to the cutsets of the original graph.

Although there are no natural analogues for plaquettes - the elementary 
2-cells - we can nevertheless define a gauge theory on an arbitrary graph
using the observation that for
a gauge theory on a regular lattice it is known that even if the action 
is formulated 
in terms of elementary loops around plaquettes, a renormalization group
transformed action will involve also other loops with the couplings for
larger loops suppressed.

The following formulation of a gauge theory on a graph does not make
reference to a specific gauge group. For each pair of neighbored
vertices $(i,j)$ define an element of the gauge group $g_{ij}$, such that
$g_{ij}=g_{ji}^{-1}$. As usual, $g_{ij}$ may be interpreted as the parallel 
transport from the representation space $W_j$ (supposed to be associated
with vertex $j$) to the isomorphic representation space $W_i$.
Consider the graph with adjacency matrix $A$ and define the matrix $A^G$ 
by
\[           (A^G)_{ij} ~=~ A_{ij} g_{ij} ~=~ \left\{ \begin{array}{ll}
           g_{ij} & \mbox{if $i$ and $j$ are neighbored}\\
             0 & \mbox{otherwise}
            \end{array} \right.     \]
$A^G$ is a linear mapping acting on the representation space 
$\hat{W}=W_1\oplus W_2\oplus \cdots W_N$, where $N$ equals the number of
vertices of the graph.
Note that $A^G$ will be hermitean in general and for the group ${\bf Z}_2$
even symmetric. The graph Laplacian in the presence of gauge fields now
reads
\[         \Delta^G ~=~ A^G - V     \, . \] 
This can be used to couple scalar fields to gauge fields on graphs.      

As before (sec.\ \ref{sec11}), powers of $A^G$ generate paths on the graph
but this time each path is ``weighted'' by its parallel transport:
\[  \left((A^G)^L\right)_{ij} ~=~ \sum_{L-{\rm paths~} j\rightarrow i}
           g_{ik_1}g_{k_1 k_2}\cdots g_{k_{L-1}j}  \, . \]
The right hand side extends over all allowed paths of length $L$ from 
vertex $j$ to vertex $i$ on the graph and for each path we obtain the
product of group elements on the links of this path, i.e.\ the parallel
transporter along this path. Taking the trace with respect to the
vertices of the graph (denoted by ``tr'') yields
\[   {\rm tr} (A^G)^L ~=~ \sum_i 
         \left( (A^G)^L \right)_{ii} ~=~ \sum_{{\rm closed~}L-{\rm paths}}
              g_{ik_1}g_{k_1 k_2}\cdots g_{k_{L-1}i} \, , \]
i.e.\ the sum over all parallel transporters along closed paths on the graph
which have length $L$ and which start and end at vertex $i$. Taking
also the trace with respect to the representation of the group (denoted
by ``Tr'') we obtain 
\[  {\rm Tr}\, {\rm tr} (A^G)^L ~=~ L \sum_{{\rm closed~}L-{\rm paths}}
            {\rm Tr}\, (g_{ik_1}g_{k_1 k_2}\cdots g_{k_{L-1}i}) \, , \]
i.e., the sum over all Wilson loops of length $L$ on the graph. (The factor
$L$ arises because each closed path occurs $L$ times on the right hand
side, as each vertex of the path can act as a starting point.)

For a gauge action on the graph we now take
\begin{equation}
\label{eqaction}
   S[A^G] ~=~ \sum_L h_L {\rm Tr}\,{\rm tr} (A^G)^L  \, , 
\end{equation}
where $h_L$ are some coupling constants which for simplicity shall only
depend on $L$.
Note that this is not the most general gauge action on a graph, which
might also include products of Wilson loops or different couplings for 
different ''shapes'' of loops (e.g., the number of links which are visited
repeatedly by the path, or the maximal distance - in the sense of sec.\ 
\ref{sec1} - between two points of the loop). Equation (\ref{eqaction})
is just the simplest choice which includes the standard Wilson action for 
regular lattices. In any case, it is expected that in a critical
limit the different choices correspond to the same universality class.    
In order to suppress contributions from very large loops (or from paths
which keep winding around some smaller loop) the coupling $h_L$ should
approach zero for $L\rightarrow \infty$ faster than a power law, e.g.\
$h_L \sim 1/L!$, or $h_L=0$ for $L$ larger than some critical length.  

Let us now return the the ${\bf Z}_2$-models on random graphs. For the
Ising model the high temperature
expansion is always an expansion in closed polygones on the graph,
representing euclidean ``world lines''. The low temperature expansion is
with respect to cutsets, and for regular lattices this corresponds to
an expansion in terms of $d-1$-dimensional closed objects (the ``world 
volumes'' of $d-2$-dimensional objects). 

For the ${\bf Z}_2$-gauge theory the situation is slightly more complicated.
The low temperature expansion on an arbitrary graph is an expansion with
respect to sets of closed loops such that an even number of loops meet at
each link. If we visualize the closed loops as the
boundaries of some surfaces (whose shape is not important and,
strictly speaking, meaningless) this would correspond to an
expansion in terms of closed surfaces.
Note, however, that our gauge action does not only contain elementary loops
even for the case of regular lattices. In principle (unless there is
a cut-off for the couplings $h_L$) all loops contribute, the larger ones
with less weight. So the closed surfaces are not only pasted together
by elementary plaquettes but also by more complicated accumulations of
plaquettes which can be interpreted as more complicated area elements.
Although this makes the expansion more complicated, the fundamental
picture of an expansion in terms of closed surfaces remains.

The low temperature expansion of the ${\bf Z}_2$-gauge theory on
an arbitrary graph may also be formulated. For each configuration of spin
variables we consider the set of loops where the curvature (product of 
parallel transports
around these loops) is $-1$. These sets satisfy certain constraints
which, on regular lattices, make them equivalent to objects of codimension
2 on the dual lattice. Again we see that the interpretation of the high
temperature expansion in terms of ``dual flux-tubes'' etc.\ only works
for regular lattices. But on the other hand it is only for these 
perturbative regimes that we expect duality to hold.

\section{Generalizations}
\label{sec4}

The following is a list of generalizations which is
far from being complete:
\begin{enumerate}
\item
The action for the random graphs may be different. The aim would be
a kind of Einstein action for which the local minima are the graph equivalents
of solutions to the Einstein equations.
\item
The gauge group can be generalized even to non-abelian cases. In sect.\
\ref{sec3} it has been shown how to couple gauge degrees of freedom to
random graphs. Matter fields, associated to the vertices of the graph,
may also be coupled to the gauge fields.
\item
Besides duality another equivalence of models is known quite well from
statistical mechanics: the equivalence between quantum statistical models
on $d$-dimensional lattices and classical statistical models on 
$d+1$-dimensional lattices. So one might wonder, if instead of classical
spin models one should consider quantum spin models on random graphs.
The construction is straight forward: For graphs with $V$ vertices
consider the $2^V$-dimensional Hilbert space ${\cal H}=\otimes_i^V {\bf C}_2$ 
and define
\[         \vec{\sigma}_i ~=~ 1 \otimes 1 \otimes \cdots 
         \otimes \vec{\sigma}\otimes \cdots \otimes 1  \, \]
to be the spin variables (Pauli matrices) attached to vertex $i$. The
Hamiltonian for the quantum Ising model on a graph with adjaceny matrix $A$
is then given by
\[           H ~=~ \mu \sum_i \sigma_i^x + \frac{1}{2}\lambda 
             \sum_{ij} A_{ij} \sigma_i^z \sigma_j^z \, . \]
In a similar way one can formulate the
quantum gauge theory Hamiltonian.
The duality relations of Wegner also hold for the quantum models (with
the dimensions of the lattices reduced by one).
It is tentative to assume a relation with spin networks  
\cite{Ashtekar} which arise in the context of a canonical quantization
prescription for gravity.
\item
Instead of graphs one might consider general cell-complexes, or, as an
intermediate step, graphs with cliques. A clique is a complete subgraph
of a graph and represents a higher dimensional simplex. For such cliques
also the spin gauge models for $p$-forms of Wegner may
be defined, because there exists a natural notion of plaquettes, volumes, 
etc. 
\end{enumerate}

It has been shown that simple models with the properties listed in the
introduction exist, but it should be stressed that these are only
``toy'' models. Other actions and gauge groups might reveal more
structures with an even closer relationship to non-perturbative string 
theories.
\vspace{0.3cm}

The author should like to acknowledge the kind hospitality at the
Albert-Einstein-Institut f\"ur Gravitationsphysik in Golm where part
of this work has been done.

\thispagestyle{empty}

\begin{thebibliography}{99}
\enlargethispage{2cm}
\bibitem{Polchinski} J.\ Polchinski, {\em String Theories}, Vol.\,1 \&
         2, Cambridge University Press, 1998.\\ 
         J.\ Polchinski, Rev.\ Mod.\ Phys.\ 68 (1996) 1245.
\bibitem{Kramers} Kramers, H.A.\ and Wannier, G.H., Phys.\ Rev.\ 60
        (1941) 252.
\bibitem{Feynman} R.\,P.\,Feynman, {\em Statistical Mechanics}, 
       Addison-Wesley Publishing Company, Inc., 1972 (reprint 1990).
\bibitem{Baxter} Baxter, R.J., {\em Exactly Solved Models in Statistical
         Mechanics}, Academic Press, 1982.
\bibitem{Wegner} F.\,J.\,Wegner, J.~Math.~Phys.\,12 (1971) 2259.
\bibitem{Coleman} S.\ Coleman, Phys.\ Rev.\ D\,11 (1975) 2088.
\bibitem{Mandelstam} S.\ Mandelstam, Phys.\ Rev.\ D\,11 (1975) 3026.
\bibitem{Olive} C.\ Montonen and D.\ Olive, 
                           Phys.\ Lett.\ 72B (1977) 117.\\
        E.\ Witten and D.\ Olive, Phys.\ Lett.\ 78B (1978) 97.\\
        H.\ Osborne, Phys.\ Lett.\ 83B (1979) 321. 
\bibitem{Maldacena} O.\ Aharony, S.S.\ Gubser, J.\ Maldacena, H.\ Ooguri,
        Y.\ Oz; Phys.\ Rep.\ 323 (2000) 183. 
\bibitem{Rothe} H.\,J.\,Rothe, {\em Lattice Gauge Theories}, World Scientific,
        Singapore, 2nd ed., 1997.
\bibitem{Biggs} N.\,Biggs, {\em Algebraic Graph Theory}, Cambridge 
       University Press, 2nd ed., 1993.
\bibitem{Stauffer} D.\ Stauffer and A.\ Aharony, {\em Introduction to 
          Percolation Theory}, Taylor and Francis, 2nd ed.\ (1994).
\bibitem{Requardt} T.\ Nowotny and M.\ Requardt, J.\ Phys.~A31 (1998) 2447.
\bibitem{Ambjorn} J.\ Ambjorn, B.\ Durhuus, J.\ Fr\"ohlich, P.\ Orlando;
        Nucl.\ Phys.~B270 (1986) 457. 
\bibitem{Regge} T.\ Regge, Nuovo Cimento 19 (1961) 558.
\bibitem{Ambjorn2} J.\ Ambjorn, {\em Quantization of Geometry}, Lectures
         presented at the 1994 Les Houches Summer School ``Fluctuating
         Geometries in Statistical Mechanics and Field Theory''.\\
        F.\ David, {\em Simplicial Quantum Gravity and Random Lattices},
         Lectures presented at the 1992 Les Houches Summer School
         ``Gravitation and Quantizations''.
\bibitem{Pauli} W.\ Pauli, {\em Theory of Relativity}, Pergamon Press,
          London, 1958.
\bibitem{Gilkey} P.B.\ Gilkey, {\em The index theorem and the heat
          equation}; Math.\ Lect.\ Series 4, Boston, Mass., Publish or
         Perish, Inc., 1974.
\bibitem{Eguchi} T.\ Eguchi, P.B.\ Gilkey, A.J.\ Hanson, {\em Gravitation,
         gauge theories and differential geometry}, Phys.\ Rep.\ 66 (1980)
         213.
\bibitem{Doob} D.M.\ Cvetkovi\'c, M.\ Doob, H.\ Sachs, {\em Spectra of
         Graphs}, 3rd ed., Johann Ambrosius Barth Verlag, Heidelber,
         Leipzig, 1995.
\bibitem{Ashtekar} A.\ Ashtekar, J.\ Lewandowski, D.\ Marolf, J.\ Mourag,
         T.\ Thiemann; J.\ Math.\ Phys.\ 36 (1995) 6456.

\end{thebibliography}
\end{document}